\documentclass[submission,creativecommons]{eptcs}
\providecommand{\event}{ACL2 2011}
\usepackage{xspace}

\title{Bit-Blasting ACL2 Theorems}

\author{
Sol Swords and Jared Davis
\institute{Centaur Technology Inc.\\
7600-C N. Capital of Texas Hwy, Suite 300\\
Austin, TX 78731} \\
\email{\{sswords,jared\}@centtech.com}
}
\def\titlerunning{Bit-Blasting ACL2 Theorems}
\def\authorrunning{Sol Swords and Jared Davis}

\begin{document}
\maketitle

\begin{abstract}

  Interactive theorem proving requires a lot of human guidance.  Proving a
  property involves (1) figuring out why it holds, then (2) coaxing the theorem
  prover into believing it.  Both steps can take a long time.  We explain how
  to use \emph{GL}, a framework for proving finite ACL2 theorems with BDD- or
  SAT-based reasoning.  This approach makes it unnecessary to deeply understand
  why a property is true, and automates the process of admitting it as a
  theorem.  We use GL at Centaur Technology to verify execution units for x86
  integer, MMX, SSE, and floating-point arithmetic.

\end{abstract}

\section{Introduction}
\label{sec:introduction}

In hardware verification you often want to show that some circuit implements its
specification.  Many of these problems are in the scope of fully automatic
decision procedures like SAT solvers.  When these tools can be used, there are
good reasons to prefer them over \emph{The Method}~\cite{00-kaufmann-car} of
traditional, interactive theorem proving.  For instance, these tools can:

\begin{itemize}

\item Reduce the level of human understanding needed in the initial
  process of developing the proof;

\item Provide clear counterexamples, whereas failed ACL2 proofs can
  often be difficult to debug; and

\item Ease the maintenance of the proof, since after the design changes
  they can often find updated proofs without help.

\end{itemize}

\emph{GL}~\cite{10-swords-dissertation} is a framework for proving
\emph{finite} ACL2 theorems---those which, at least in principle, could be
established by exhaustive testing---by bit-blasting with a Binary Decision
Diagram (BDD) package or a SAT solver.  These approaches have much higher
capacity than exhaustive testing.  We are using GL heavily at Centaur
Technology~\cite{11-slobodova-framework,10-hardin-centaur,09-hunt-fadd}.  So
far, we have used it to verify RTL implementations of floating-point addition,
multiplication, and conversion operations, as well as hundreds of bitwise and
arithmetic operations on scalar and packed integers.

This paper is an introduction to GL and a practical guide for using it to prove
ACL2 theorems.  For a comprehensive treatment of the implementation of GL, see
Swords' dissertation~\cite{10-swords-dissertation}.  Additional details about
particular commands can be found in the online documentation with \texttt{:doc
  gl}.

GL is the successor of Boyer and Hunt's~\cite{09-boyer-g} \emph{G} system
(Section \ref{sec:related}), and its name stands for \emph{G in the Logic}.
The G system was written as a raw Lisp extension of the ACL2 kernel, so using
it meant trusting this additional code.  In contrast, GL is implemented as ACL2
books and its proof procedure is formally verified by ACL2, so the only code we
have to trust besides ACL2 is the ACL2(h) extension that provides hash-consing
and memoization~\cite{06-boyer-acl2h}.  Like the G system, GL can prove
theorems about ordinary ACL2 definitions; you are not restricted to some small
subset of the language.

How does GL work?  You can probably imagine writing a bit-based encoding of
ACL2 objects.  For instance, you might represent an integer with some structure
that contains a 2's-complement list of bits.  GL uses an encoding like this,
except that Boolean expressions take the place of the bits.  We call these
structures \emph{symbolic objects} (Section \ref{sec:symbolic-objects}).

GL provides a way to effectively compute with symbolic objects; e.g., it can
``add'' two integers whose bits are expressions, producing a new symbolic
object that represents their sum.  GL can perform similar computations for most
ACL2 primitives.  Building on this capability, it can \emph{symbolically
  execute} terms (Section \ref{sec:symbolic-execution}). The result of a
symbolic execution is a new symbolic object that captures all the possible
values the result could take.

Symbolic execution can be used as a proof procedure (Section
\ref{sec:proving-theorems}).  To prove a theorem, we first symbolically execute
its goal formula, then show the resulting symbolic object cannot represent
\texttt{nil}.  GL provides a \texttt{def-gl-thm} command that makes it easy to
prove theorems with this approach (Section \ref{sec:def-gl-thm}).  It handles
all the details of working with symbolic objects, and only needs to be told how
to represent the variables in the formula.

Like any automatic procedure, GL has a certain capacity.  But when these limits
are reached, you may be able to increase its capacity by:

\begin{itemize}
\item Optimizing its symbolic execution strategy to use more efficient
  definitions (Section \ref{sec:optimization}),

\item Decomposing difficult problems into easier subgoals using an automatic tool
  (Section \ref{sec:def-gl-param-thm}), or

\item Using a SAT backend (Section \ref{sec:aig-mode}) that outperforms BDDs
  on some problems.
\end{itemize}
There are also some good tools and techniques for debugging failed proofs
(Section \ref{sec:debugging}).

\subsection{Example: Counting Bits}
\label{sec:counting-bits}

Let's use GL to prove a theorem.  The following C code, from Anderson's
\emph{Bit Twid\-dl\-ing Hacks}~\cite{11-anderson-bit-hacks} page, is a fast way
to count how many bits are set in a 32-bit integer.

\[
\begin{array}{l}
\texttt{v = v - ((v >> 1) \& 0x55555555);} \\
\texttt{v = (v \& 0x33333333) + ((v >> 2) \& 0x33333333);} \\
\texttt{c = ((v + (v >> 4) \& 0xF0F0F0F) * 0x1010101) >> 24;} \\
\end{array}
\]

We can model this in ACL2 as follows.  It turns out that using
arbitrary-precision addition and subtraction does not affect the result, but we
must take care to use a 32-bit multiply to match the C code.

\[
\begin{array}{l}
\texttt{(defun 32* (x y)} \\
\texttt{~~(logand (* x y) (1- (expt 2 32))))} \\
\texttt{} \\
\texttt{(defun fast-logcount-32 (v)} \\
\texttt{~~(let* ((v (- v (logand (ash v -1) \#x55555555)))} \\
\texttt{~~~~~~~~~(v (+ (logand v \#x33333333) (logand (ash v -2) \#x33333333))))} \\
\texttt{~~~~(ash (32* (logand (+ v (ash v -4)) \#xF0F0F0F) \#x1010101) -24)))} \\
\end{array}
\]

We can then use GL to prove \texttt{fast-logcount-32} computes the same result
as ACL2's built-in \texttt{logcount} function for all unsigned 32-bit inputs.

\[
\begin{array}{l}
\texttt{(def-gl-thm fast-logcount-32-correct} \\
\texttt{~~:hyp~~~(unsigned-byte-p 32 x)} \\
\texttt{~~:concl (equal (fast-logcount-32 x)} \\
\texttt{~~~~~~~~~~~~~~~~(logcount x))} \\
\texttt{~~:g-bindings `((x ,(g-int 0 1 33))))} \\
\end{array}
\]

The \texttt{:g-bindings} form is the only help GL needs from the user.
It tells GL how to construct a symbolic object that can represent every value
for \texttt{x} that satisfies the hypothesis (we explain what it means in
later sections).  No arithmetic books or lemmas are required---we actually don't
even know why this algorithm works.  The proof completes in 0.09 seconds and
results in the following ACL2 theorem.

\[
\begin{array}{l}
\texttt{(defthm fast-logcount-32-correct} \\
\texttt{~~(implies (unsigned-byte-p 32 x)} \\
\texttt{~~~~~~~~~~~(equal (fast-logcount-32 x)} \\
\texttt{~~~~~~~~~~~~~~~~~~(logcount x)))} \\
\texttt{~~:hints ((gl-hint ...)))} \\
\end{array}
\]

Why not just use exhaustive testing?  We wrote a fixnum-optimized
exhaustive-testing function that can cover the $2^{32}$ cases in 143 seconds.
This is slower than GL but still seems reasonable.  On the other hand,
exhaustive testing is clearly incapable of scaling to the 64-bit and 128-bit
versions of this algorithm, whereas GL completes the proofs in 0.18 and 0.58
seconds, respectively.

Like exhaustive testing, GL can generate counterexamples to non-theorems.  At
first, we didn't realize we needed to use a 32-bit multiply in
\texttt{fast-logcount-32}, and we just used an arbitrary-precision multiply
instead.  The function still worked for test cases like \texttt{0}, \texttt{1},
\texttt{\#b111}, and \texttt{\#b10111}, but when we tried to prove its
correctness, GL showed us three counterexamples, \texttt{\#x80000000},
\texttt{\#xFFFFFFFF}, and \texttt{\#x9448C263}.  By default, GL generates a
first counterexample by setting bits to 0 wherever possible, a second by
setting bits to 1, and a third with random bit settings.

\subsection{Example: UTF-8 Decoding}
\label{sec:utf-8}

Davis~\cite{06-davis-input} used exhaustive testing to prove lemmas toward the
correctness of UTF-8 processing functions.  The most difficult proof carried
out this way was a well-formedness and inversion property for four-byte
UTF-8 sequences, which involved checking $2^{32}$ cases.  Davis' proof takes 67
seconds on our computer.  It involves four testing functions and five lemmas
about them; all of this is straightforward but mundane.  The testing functions
are guard-verified and optimized with \texttt{mbe} and type declarations for
better performance.

We used GL to prove the same property.  The proof (included in the supporting
materials) completes in 0.17 seconds and requires no testing functions or
supporting lemmas.

% \fixverbatim
% \begin{verbatim}
% (def-gl-thm lemma-5-by-gl
%  :hyp (utf8-combine4-guard x1 x2 x3 x4)
%  :concl
%  (and
%   (uchar? (utf8-combine4 x1 x2 x3 x4))
%   (utf8-table35-ok? (utf8-combine4 x1 x2 x3 x4)
%                     (list x1 x2 x3 x4))
%   (utf8-table36-ok? (utf8-combine4 x1 x2 x3 x4)
%                     (list x1 x2 x3 x4))
%   (equal (uchar=>utf8 (utf8-combine4 x1 x2 x3 x4))
%          (list x1 x2 x3 x4)))
%  :rule-classes nil
%  :g-bindings `((x1 ,(g-int 0 1 9))
%                (x2 ,(g-int 9 1 9))
%                (x3 ,(g-int 18 1 9))
%                (x4 ,(g-int 27 1 9))))
% \end{verbatim}
% \fixverbatim

\subsection{Getting GL}

GL is included in ACL2 4.3, and the development version is available from the
ACL2 Books repository, \url{http://acl2-books.googlecode.com/}.  Note that
using GL requires ACL2(h), which is best supported on 64-bit Clozure Common
Lisp.  BDD operations can be memory intensive, so we recommend using a computer
with at least 8 GB of memory.  Instructions for building GL can be found in
\texttt{centaur/README}, and it can be loaded with
\[
\texttt{(include-book "centaur/gl/gl" :dir :system)}.
\]

\section{GL Basics}
\label{sec:gl-basics}

At its heart, GL works by manipulating Boolean expressions.  There are many
ways to represent Boolean expressions.  GL currently supports a hons-based BDD
package~\cite{06-boyer-acl2h} and also has support for using a hons-based
And-Inverter Graph (AIG) representation with an external SAT solver.

For any particular proof, the user can choose to work in \emph{BDD mode} (the
default) or \emph{AIG mode}.  Each representation has strengths and weaknesses,
and the choice of representation can significantly impact performance.  We give
some advice about choosing proof modes in Section \ref{sec:aig-mode}.

\newcommand{\Btrue}{\ensuremath{\mathit{true}}\xspace}
\newcommand{\Bfalse}{\ensuremath{\mathit{false}}\xspace}

The GL user does not need to know how BDDs and AIGs are represented; in this
paper we just adopt a conventional mathematical syntax to describe Boolean
expressions, e.g., $\Btrue$, $\Bfalse$, $A \wedge B$, $\neg C$, etc.

\subsection{Symbolic Objects}
\label{sec:symbolic-objects}
\newcommand{\SP}{\texttt{ }}

GL groups Boolean expressions into  \emph{symbolic objects}.  Much like
a Boolean expression can be evaluated to obtain a Boolean value, a symbolic
object can be evaluated to produce an ACL2 object.  There are several kinds of
symbolic objects, but numbers are a good start.  GL represents symbolic, signed
integers as
\[
\texttt{(:g-number~$\mathit{lsb\textrm{-}bits}$)},
\]
where \emph{lsb-bits} is a list of Boolean expressions that represent the two's
complement bits of the number.  The bits are in lsb-first order, and the last,
most significant bit is the sign bit.  For instance, if $p$ is the following
\texttt{:g-number},
\[
p = \texttt{(:g-number (}\Btrue \SP \Bfalse \SP A \wedge B \SP \Bfalse \texttt{))},
\]
then $p$ represents a 4-bit, signed integer whose value is either 1 or 5,
depending on the value of $A \wedge B$.

GL uses another kind of symbolic object to represent ACL2 Booleans.  In particular,
\[
\texttt{(:g-boolean~.~$\mathit{val}$)}
\]
represents \texttt{t} or \texttt{nil} depending on the Boolean expression
\emph{val}.  For example,
\[
\texttt{(:g-boolean~.~$\neg(A \wedge B)$)}
\]
is a symbolic object whose value is \texttt{t} when $p$ has value 1, and
\texttt{nil} when $p$ has value 5.

GL has a few other kinds of symbolic objects that are also tagged with
keywords, such as \texttt{:g-var} and \texttt{:g-apply}.  But an ACL2 object
that does not have any of these special keywords within it is \emph{also}
considered to be a symbolic object, and just represents itself.  Furthermore, a
cons of two symbolic objects represents the cons of the two objects they
represent.  For instance,
\[
\texttt{(1~.~(:g-boolean~.~$A \wedge B$))}
\]
represents either \texttt{(1~.~t)} or \texttt{(1~.~nil)}.  Together, these
conventions allow GL to avoid lots of tagging as symbolic objects are
manipulated.

\newcommand{\Stest}{\ensuremath{\mathit{test}}}
\newcommand{\Sthen}{\ensuremath{\mathit{then}}}
\newcommand{\Selse}{\ensuremath{\mathit{else}}}

One last kind of symbolic object we will mention represents an if-then-else
among other symbolic objects.  Its syntax is
\[
\texttt{(:g-ite~$\Stest$~$\Sthen$~.~$\Selse$)},
\]
where $\Stest$, $\Sthen$, and $\Selse$ are themselves symbolic objects.  The
value of a \texttt{:g-ite} is either the value of $\Sthen$ or of $\Selse$,
depending on the value of $\Stest$.  For example,
\[
\begin{array}{l}
\texttt{(:g-ite~(:g-boolean~.~$A$)} \\
\texttt{~~~~~~~~(:g-number~($B$~$A$~\Bfalse))} \\
\texttt{~~~~~~~~.~\#\textbackslash{}C)}
\end{array}
\]
represents either 2, 3, or the character \texttt{C}.

GL doesn't have a special symbolic object format for ACL2 objects other than
numbers and Booleans.  But it is still possible to create symbolic objects that
take any finite range of values among ACL2 objects, by using a nesting of
\texttt{:g-ite}s where the tests are \texttt{:g-boolean}s.

\subsection{Computing with Symbolic Objects}
\label{sec:symbolic-execution}

Once we have a representation for symbolic objects, we can perform symbolic
executions on those objects.  For instance, recall the symbolic number $p$
which can have value 1 or 5,
\[
p = \texttt{(:g-number (}\Btrue \SP \Bfalse \SP A \wedge B \SP \Bfalse \texttt{))}.
\]
We might symbolically add 1 to $p$ to obtain a new symbolic number, say $q$,
\[
q = \texttt{(:g-number (}\Bfalse \SP \Btrue \SP A \wedge B \SP \Bfalse \texttt{))},
\]
which represents either 2 or 6.  Suppose $r$ is another symbolic number,
\[
r = \texttt{(:g-number (}A \SP \Bfalse \SP \Btrue \SP \Bfalse \texttt{))},
\]
which represents either 4 or 5.  We might add $q$ and $r$ to obtain $s$,
\[
s = \texttt{(:g-number (}A \SP \Btrue \SP \neg(A \wedge B) \SP A \wedge B \SP
\Bfalse \texttt{))},
\]
whose value can be 6, 7, or 11.  Why can't $s$ be 10 if $q$ can be 6 and $r$
can be 4?  This combination isn't possible because $q$ and $r$ involve the same
expression, $A$.  The only way for $r$ to be 4 is for $A$ to be false, but then
$q$ must be 2.

The underlying algorithm GL uses for symbolic additions is just a ripple-carry
addition on the Boolean expressions making up the bits of the two numbers.
Performing a symbolic addition, then, means constructing new
BDDs or AIGs, depending on which mode is being used.

GL has built-in support for symbolically executing most ACL2 primitives.
Generally, this is done by cases on the types of the symbolic objects being
passed in as arguments.  For instance, if we want to symbolically execute
\texttt{consp} on $s$, then we are asking whether a \texttt{:g-number} may ever
represent a cons, so the answer is simply \texttt{nil}.  Similarly, if we ever
try to add a \texttt{:g-boolean} to a \texttt{:g-number}, by the ACL2 axioms
the \texttt{:g-boolean} is simply treated as 0.

Beyond these primitives, GL provides what is essentially a McCarthy-style
interpreter~\cite{60-mccarthy-recursive} for symbolically executing terms.  By
default, it expands function definitions until it reaches primitives, with some
special handling for \texttt{if}.  For better performance, its
interpretation scheme can be customized with more efficient definitions and
other optimizations, as described in Section \ref{sec:optimization}.

%% GL provides symbolic way to represent any ACL2 object, and symbolic
%% implementations of almost all ACL2 primitives that allow these sorts of
%% operations to be carried out.  Using these primitives, it can symbolically
%% interpret user-defined functions and the formulas that occur in theorems.

\subsection{Proving Theorems by Symbolic Execution}
\label{sec:proving-theorems}

\newcommand{\Xbest}{\ensuremath{x_{\mathit{best}}}\xspace}
\newcommand{\Xinit}{\ensuremath{x_{\mathit{init}}}\xspace}

To see how symbolic execution can be used to prove theorems, let's return to
the bit-counting example, where our goal was to prove
\[
\begin{array}{l}
\texttt{(implies (unsigned-byte-p 32 x)} \\
\texttt{~~~~~~~~~(equal (fast-logcount-32 x)} \\
\texttt{~~~~~~~~~~~~~~~~(logcount x)))}. \\
\end{array}
\]
The basic idea is to first symbolically execute the above formula, and then
check whether it can ever evaluate to \texttt{nil}.  But to do this symbolic
execution, we need some symbolic object to represent \texttt{x}.

We want our symbolic execution to cover all the cases necessary for proving the
theorem, namely all \texttt{x} for which the hypothesis
\texttt{(unsigned-byte-p 32 x)} holds.  In other words, the symbolic object we
choose needs to be able to represent any integer from 0 to $2^{32}-1$.

Many symbolic objects cover this range.  As notation, let $b_0,b_1,\dots$
represent independent Boolean variables in our Boolean expression
representation.  Then, one suitable object is:
\[
\texttt{(:g-number ($b_0$~$b_1$~$\dots$~$b_{31}$~$b_{32}$))}.
\]
Why does this have 33 variables?  The final bit, $b_{32}$, represents the sign,
so this object covers the integers from $-2^{32}$ to $2^{32}-1$.  We could
instead use a 34-bit integer, or a 35-bit integer, or some esoteric creation
involving \texttt{:g-ite} forms.  But perhaps the best object to use would be:
\[
\Xbest = \texttt{(:g-number ($b_0$~$b_1$~$\dots$~$b_{31}$~$\Bfalse$))},
\]
since it covers exactly the desired range using the simplest possible Boolean
expressions.

Suppose we choose \Xbest to stand for \texttt{x}.  We can now
symbolically execute the goal formula on that object.

What does this involve?  First, \texttt{(unsigned-byte-p 32 x)} produces the
symbolic result \texttt{t}, since it is always true of the possible values of
\Xbest.  It would have been equally valid for this to produce
\texttt{(:g-boolean~.~$\Btrue$)}, but GL prefers to produce constants
when possible.

Next, the \texttt{(fast-logcount-32 x)} and \texttt{(logcount x)} forms each
yield \texttt{:g-number} objects whose bits are Boolean
expressions in the variables $b_0, \dots, b_{31}$.  For example, the least
significant bit will be an expression representing the XOR of all these
variables.

Finally, we symbolically execute \texttt{equal} on these two results.  This
compares the Boolean expressions for their bits to determine if they are
equivalent, and produces a symbolic object representing the answer.

So far we have basically ignored the differences between using BDDs and AIGs as
our Boolean expression representation.  But here, the two approaches produce
very different answers:

\begin{itemize}
\item Since BDDs are canonical, the expressions for the bits of the two numbers
  are syntactically equal, and the result from \texttt{equal} is simply \texttt{t}.

\item With AIGs, the expressions for the bits are semantically equivalent but
  not syntactically equal.  The result is therefore
  \texttt{(:g-boolean~.~$\phi$)}, where $\phi$ is a large Boolean expression in
  the variables $b_0, \dots, b_{31}$.  The fact that $\phi$ always evaluates to
  \Btrue is not obvious just from its syntax.
\end{itemize}

At this point we have completed the symbolic execution of our goal formula,
obtaining either \texttt{t} in BDD mode, or this \texttt{:g-boolean} object in
AIG mode.  Recall that to prove theorems using symbolic execution, the idea is
to symbolically execute the goal formula and then check whether its symbolic
result can represent \texttt{nil}.  If we are using BDDs, it is obvious that
\texttt{t} cannot represent \texttt{nil}.  With AIGs, we simply ask a SAT
solver whether $\phi$ can evaluate to \Bfalse, and find that it cannot.  This
completes the proof.

GL automates this proof strategy, taking care of many of the details relating
to creating symbolic objects, ensuring that they cover all the possible cases,
and ensuring that \texttt{nil} cannot be represented by the symbolic result.
When GL is asked to prove a non-theorem, it can generate counterexamples by
finding assignments to the Boolean variables that cause the result to become
\texttt{nil}.

\section{Using DEF-GL-THM}
\label{sec:def-gl-thm}

The \texttt{def-gl-thm} command is the main interface for using GL to prove
theorems.  Here is the command we used in the bit-counting example.

\[
\begin{array}{l}
\texttt{(def-gl-thm fast-logcount-32-correct} \\
\texttt{~~:hyp~~~(unsigned-byte-p 32 x)} \\
\texttt{~~:concl (equal (fast-logcount-32 x)} \\
\texttt{~~~~~~~~~~~~~~~~(logcount x))} \\
\texttt{~~:g-bindings `((x ,(g-int 0 1 33))))} \\
\end{array}
\]

Unlike an ordinary \texttt{defthm} command, \texttt{def-gl-thm} takes separate
hypothesis and conclusion terms (its \texttt{:hyp} and \texttt{:concl}
arguments).  This separation allows GL to use the hypothesis to limit the scope
of the symbolic execution it will perform.  The user must also provide GL with
\texttt{:g-bindings} that describe the symbolic objects to use for each free
variable in the theorem (Section \ref{sec:writing-g-bindings}).

What are these bindings?  In the \texttt{fast-logcount-32-corr\-ect} theorem, we
used a convenient function, \texttt{g-int}, to construct the
\texttt{:g-bindings}.  Expanding this away, here are the actual bindings:
\[
\texttt{((x (:g-number (0 1 2 $\dots$ 32))))}.
\]
The \texttt{:g-bindings} argument uses a slight modification of the symbolic
object format where the Boolean expressions are replaced by distinct
natural numbers, each representing a Boolean variable.  In this
case, our binding for \texttt{x} stands for the following symbolic object:
\[
\Xinit = \texttt{(:g-number ($b_0$~$b_1$~$\dots$~$b_{31}$~$b_{32}$))}.
\]
Note that \Xinit is not the same object as \Xbest from Section
\ref{sec:proving-theorems}---its sign bit is $b_{32}$ instead of \Bfalse, so
\Xinit can represent any 33-bit signed integer whereas \Xbest only represents
32-bit unsigned values.  In fact, the \texttt{:g-bindings} syntax does not even
allow us to describe objects like \Xbest, which has the constant \Bfalse
instead of a variable as one of its bits.

There is a good reason for this restriction.  One of the steps in our proof
strategy is to prove \emph{coverage}: we need to show the symbolic objects we
are starting out with have a sufficient range of values to cover all cases for
which the hypothesis holds (Section \ref{sec:proving-coverage}).  The
restricted syntax permitted by \texttt{:g-bindings} ensures that the range of
values represented by each symbolic object is easy to determine.  Because of
this, coverage proofs are usually automatic.

Despite these restrictions, GL will still end up using \Xbest to carry out the
symbolic execution.  GL optimizes the original symbolic objects inferred from
the \texttt{:g-bindings} by using the hypothesis to reduce the space of objects
that are represented.  In BDD mode this optimization uses \emph{BDD
  parametrization}~\cite{99-aagaard-param}, which restricts the symbolic
objects so they cover exactly the inputs recognized by the hypothesis.  In AIG
mode we use a lighter-weight transformation that replaces variables with
constants when the hypothesis sufficiently restricts them.  In this example,
either optimization transforms \Xinit into \Xbest.

\subsection{Writing G-Bindings Forms}
\label{sec:writing-g-bindings}

In a typical \texttt{def-gl-thm} command, the \texttt{:g-bindings} should have
an entry for every free variable in the theorem.  Here is an example that shows
some typical bindings.

\[
\begin{array}{l}
\texttt{:g-bindings~'((flag~~~(:g-boolean~.~0))} \\
\texttt{~~~~~~~~~~~~~~(a-bus~~(:g-number~(1~3~5~7~9)))} \\
\texttt{~~~~~~~~~~~~~~(b-bus~~(:g-number~(2~4~6~8~10)))} \\
\texttt{~~~~~~~~~~~~~~(mode~~~(:g-ite~(:g-boolean~.~11)~exact~.~fast))} \\
\texttt{~~~~~~~~~~~~~~(opcode~\#b0010100))} \\
\end{array}
\]

These bindings allow \texttt{flag} to take an arbitrary Boolean value,
\texttt{a-bus} and \texttt{b-bus} any five-bit signed integer values,
\texttt{mode} either the symbol \texttt{exact} or \texttt{fast}, and
\texttt{opcode} only the value 20.\footnote{Note that since \texttt{\#b0010100}
  is not within a \texttt{:g-boolean} or \texttt{:g-number} form, it is
  \emph{not} the index of a Boolean variable.  Instead, like the symbols
  \texttt{exact} and \texttt{fast}, it is just an ordinary ACL2 constant that
  stands for itself, i.e., 20.}

% Advantages to the footnote above:
% - it talks about indices which we haven't covered yet, so better to push it down
% - it is too long to put inline and is really a trivial sort of thing

Within \texttt{:g-boolean} and \texttt{:g-number} forms, natural number indices
take the places of Boolean expressions.  The indices used throughout all of the
bindings must be distinct, and represent free, independent Boolean variables.
In BDD mode these indices have additional meaning: they specify the BDD
variable ordering, with smaller indices coming first in the order.  This
ordering can greatly affect performance.  In AIG mode the choice of indices has
no particular bearing on efficiency.

How do you choose a good BDD ordering?  It is often good to interleave the bits
of data buses that are going to be combined in some way.  It is also typically
a good idea to put any important control signals such as opcodes and mode
settings before the data buses.

Often the same \texttt{:g-bindings} can be used throughout several theorems,
either verbatim or with only small changes.  In practice, we almost always
generate the \texttt{:g-bindings} forms by calling functions or macros.  One
convenient function is
\[
\texttt{(g-int start by n)},
\]
which generates a \texttt{:g-number} form with \texttt{n} bits, using
indices that start at \texttt{start} and increment by \texttt{by}.  This is
particularly useful for interleaving the bits of numbers, as we did for the
\texttt{a-bus} and \texttt{b-bus} bindings above:
\[
\begin{array}{l}
\texttt{(g-int 1 2 5)} \rightarrow \texttt{(:g-number (1 3 5 7 9))} \\
\texttt{(g-int 2 2 5)} \rightarrow \texttt{(:g-number (2 4 6 8 10))}.
\end{array}
\]

\subsection{Proving Coverage}
\label{sec:proving-coverage}

There are really two parts to any GL theorem.  First, we need to symbolically
execute the goal formula and ensure it cannot evaluate to \texttt{nil}.  But
in addition to this, we must ensure that the objects we use to represent the
variables of the theorem cover all the cases that satisfy the hypothesis.  This
part of the proof is called the \emph{coverage obligation}.

For \texttt{fast-logcount-32-correct}, the coverage obligation is to
show that our binding for \texttt{x} is able to represent every integer
from 0 to $2^{32}-1$.  This is true of \Xinit, and the coverage proof goes
through automatically.

But suppose we forget that \texttt{:g-number}s use a signed representation, and
attempt to prove \texttt{fast-log\-count-32-correct} using the following
(incorrect) g-bindings.
\[
\texttt{:g-bindings `((x ,(g-int 0 1 32)))}
\]
This looks like a 32-bit integer, but because of the sign bit it does not cover
the intended unsigned range.  If we submit the \texttt{def-gl-thm} command
with these bindings, the symbolic execution part of the proof is still successful.
But this execution has only really shown the goal holds for 31-bit unsigned
integers, so \texttt{def-gl-thm} prints the message
\[
\texttt{ERROR: Coverage proof appears to have failed.}
\]
and leaves us with a failed subgoal,
\[
\begin{array}{l}
\texttt{(implies (and (integerp x)} \\
\texttt{~~~~~~~~~~~~~~(<= 0 x)} \\
\texttt{~~~~~~~~~~~~~~(< x 4294967296))} \\
\texttt{~~~~~~~~~(< x 2147483648))}. \\
\end{array}
\]
This goal is clearly not provable: we are trying to show \texttt{x} must be
less than $2^{31}$ (from our \texttt{:g-bindings}) whenever
it is less than $2^{32}$ (from the hypothesis).

Usually when the \texttt{:g-bindings} are correct, the coverage proof will be
automatic, so if you see that a coverage proof has failed, the first thing to
do is check whether your bindings are really sufficient.

On the other hand, proving coverage is undecidable in principle, so sometimes
GL will fail to prove coverage even though the bindings are appropriate.  For
these cases, there are some keyword arguments to \texttt{def-gl-thm} that may
help coverage proofs succeed.

First, as a practical matter, GL does the symbolic execution part of the proof
\emph{before} trying to prove coverage.  This can get in the way of debugging
coverage proofs when the symbolic execution takes a long time.  You can use
\texttt{:test-side-goals t} to have GL skip the symbolic execution and go
straight to the coverage proof.  Of course, no \texttt{defthm} is
produced when this option is used.

By default, our coverage proof strategy uses a restricted set of rules and
ignores the current theory.  It heuristically expands functions in the
hypothesis and throws away terms that seem irrelevant.  When this
strategy fails, it is usually for one of two reasons.

1. The heuristics expand too many terms and overwhelm ACL2.  GL tries to avoid
this by throwing away irrelevant terms, but sometimes this approach is
insufficient.  It may be helpful to disable the expansion of functions that are not
important for proving coverage.  The \texttt{:do-not-expand} argument allows
you to list functions that should not be expanded.

2. The heuristics throw away a necessary hypothesis, leading to unprovable
goals.  GL's coverage proof strategy tries to show that the binding for each
variable is sufficient, one variable at a time.  During this process it throws
away hypotheses that do not mention the variable, but in some cases this can be
inappropriate.  For instance, suppose the following is a coverage goal for
\texttt{b}:
\[
\begin{array}{l}
\texttt{(implies (and (natp a)} \\
\texttt{~~~~~~~~~~~~~~(natp b)} \\
\texttt{~~~~~~~~~~~~~~(< a (expt 2 15))} \\
\texttt{~~~~~~~~~~~~~~(< b a))} \\
\texttt{~~~~~~~~~(< b (expt 2 15))}.
\end{array}
\]
Here, throwing away the terms that don't mention \texttt{b} will cause the proof
to fail.  A good way to avoid this problem is to separate type and size
hypotheses from more complicated assumptions that are not important for proving
coverage, along these lines:

\[
\begin{array}{l}
\texttt{(def-gl-thm~my-theorem} \\
\texttt{~~:hyp~(and~(type-assms-1~x)} \\
\texttt{~~~~~~~~~~~~(type-assms-2~y)} \\
\texttt{~~~~~~~~~~~~(type-assms-3~z)} \\
\texttt{~~~~~~~~~~~~(complicated-non-type-assms~x~y~z))} \\
\texttt{~~:concl~...} \\
\texttt{~~:g-bindings~...} \\
\texttt{~~:do-not-expand~'(complicated-non-type-assms))}.
\end{array}
\]

For more control, you can also use the \texttt{:cov-theory-add} argument to
enable additional rules during the coverage proof, e.g.,
\texttt{:cov-theory-add '(type-rule1 type-rule2)}.

% As a heuristic, we think that if a hypothesis doesn't mention a particular
% variable, it is usually not important to that variable's coverage, and we
% can throw it away if we are concentrating on that particular variable.

% For instance, a typical :hyp for a GL theorem might be the following:

% (and (natp a)
%      (natp b)
%      (< a (expt 2 16))
%      (< b (expt 2 32)))

% Here, when we are trying to prove that the binding for B is sufficient, we
% will throw away the hyps for (natp a) and (< a (expt 2 16)) since they do
% not mention B, which simplifies the problem and leaves us with exactly the
% useful hypotheses.

% But this heuristic is sometimes inappropriate.  For instance, if we had the
% hypothesis:

% general advice for writing g-bindings, and variable orderings, etc.

\section{Optimizing Symbolic Execution}
\label{sec:optimization}

The scope of theorems GL can handle is directly impacted by its symbolic
execution performance.  It is actually quite easy to customize the way
certain terms are interpreted, and this can sometimes provide
important speedups.

GL's symbolic interpreter operates much like a basic Lisp interpreter.  To
symbolically interpret a function call, GL first eagerly interprets its
arguments to obtain symbolic objects for the actuals.  Then GL symbolically
executes the function in one of three ways:

\begin{itemize}

\item As a special case, if the actuals evaluate to concrete objects, then GL
  may be able to stop symbolically executing and just call the actual ACL2
  function on these arguments (Section \ref{sec:concrete-execution}).

\item For primitive ACL2 functions like \texttt{+}, \texttt{consp},
  \texttt{equal}, and for some defined functions like \texttt{logand} and
  \texttt{ash} where performance is important, GL uses hand-written
  functions called \emph{symbolic counterparts} that can operate on symbolic
  objects.  The advanced GL user can write new symbolic counterparts
  (Section \ref{sec:custom-symbolic-counterparts}) to speed up symbolic
  execution.

\item Otherwise, GL looks up the definition of the function, and recursively
  interprets its body in a new environment binding the formals to the symbolic
  actuals.  The way a function is written can impact its symbolic execution
  performance (Section \ref{sec:redundant-recursion}).  It is easy to instruct
  GL to use more efficient definitions for particular functions (Section
  \ref{sec:preferred-definitions}).
\end{itemize}

GL symbolically executes functions strictly according to the ACL2 logic and
does not consider guards.  An important consequence is that when \texttt{mbe}
is used, GL's interpreter follows the \texttt{:logic} definition instead of the
\texttt{:exec} definition, since it might be unsound to use the \texttt{:exec}
version of a definition without establishing the guard is met.  Also, while GL
can symbolically simulate functions that take user-defined stobjs or even the
ACL2 \texttt{state}, it does not operate on ``real'' stobjs; instead, it uses
the logical definitions of the relevant stobj operations, which do not provide
the performance benefits of destructive operations.
Non-executable functions cannot be symbolically executed.

\subsection{Avoiding Redundant Recursion}
\label{sec:redundant-recursion}

Here are two ways to write a list-filtering function.

\[
\begin{array}{l}
\texttt{(defun~filter1~(x)} \\
\texttt{~~(cond~((atom~x)} \\
\texttt{~~~~~~~~~nil)} \\
\texttt{~~~~~~~~((element-okp~(car~x))~~~~~~~~~~~~~~~;;~keep~it} \\
\texttt{~~~~~~~~~(cons~(car~x)~(filter1~(cdr~x))))} \\
\texttt{~~~~~~~~(t~~~~~~~~~~~~~~~~~~~~~~~~~~~~~~~~~~~;;~skip~it} \\
\texttt{~~~~~~~~~(filter1~(cdr~x)))))} \\
\end{array}
\]

This definition can be inefficient for symbolic execution.  Suppose we are
symbolically executing \texttt{filter1}, and the \texttt{element-okp} check has
produced a symbolic object that can take both \texttt{nil} and non-\texttt{nil}
values.  Then, we proceed by symbolically executing both the keep- and
skip-branches, and construct a \texttt{:g-ite} form for the result.  Since we
have to evaluate the recursive call twice, this execution becomes exponential
in the length of \texttt{x}.

We can avoid this blow-up by consolidating the recursive calls, as follows.

\[
\begin{array}{l}
\texttt{(defun~filter2~(x)} \\
\texttt{~~(if~(atom~x)} \\
\texttt{~~~~~~nil} \\
\texttt{~~~~(let~((rest~(filter2~(cdr~x))))} \\
\texttt{~~~~~~(if~(element-okp~(car~x))} \\
\texttt{~~~~~~~~~~(cons~(car~x)~rest)} \\
\texttt{~~~~~~~~rest))))} \\
\end{array}
\]

This is not a novel observation; Reeber~\cite{07-reeber-dissertation} suggests
the same sort of optimization for unrolling recursive functions in SULFA.

Of course, \texttt{filter1} is probably slightly better for concrete execution
since it has a tail call in at least some cases.  If we do not want to change
the definition of \texttt{filter1}, we can simply tell GL to use the
\texttt{filter2} definition instead, as described in the next section.  We
currently do not try to automatically apply this kind of optimization, though
we may explore this in future work.

\subsection{Preferred Definitions}
\label{sec:preferred-definitions}

To instruct GL to symbolically execute \texttt{filter2} in place of \texttt{filter1},
we can do the following:

\[
\begin{array}{l}
\texttt{(defthm~filter1-for-gl} \\
\texttt{~~(equal~(filter1~x)~(filter2~x))} \\
\texttt{~~:rule-classes~nil)} \\
\texttt{} \\
\texttt{(gl::set-preferred-def~filter1~filter1-for-gl)} \\
\end{array}
\]

The \texttt{gl::set-preferred-def} form extends a table that GL consults when
expanding a function's definition.  Each entry in the table pairs a function
name with the name of a theorem.  The theorem must state that a call of the
function is unconditionally equal to some other term.  When GL encounters a
call of a function in this table, it replaces the call with the right-hand
side of the theorem, which is justified by the theorem.  So after the above
event, GL will replace calls of \texttt{filter1} with \texttt{filter2}.

As another example of a preferred definition, GL automatically optimizes the
definition of \texttt{evenp}, which ACL2 defines as follows:
\[
\texttt{(evenp x)} = \texttt{(integerp (* x (/ 2)))}.
\]
This definition is basically unworkable since GL provides little support for
rational numbers.  However, GL has an efficient, built-in implementation of
\texttt{logbitp}.  So to permit the efficient execution of \texttt{evenp}, GL
proves the following identity and uses it as \texttt{evenp}'s preferred
definition.

\[
\begin{array}{l}
\texttt{(defthm~evenp-is-logbitp} \\
\texttt{~~(equal~(evenp~x)} \\
\texttt{~~~~~~~~~(or~(not~(acl2-numberp~x))} \\
\texttt{~~~~~~~~~~~~~(and~(integerp~x)} \\
\texttt{~~~~~~~~~~~~~~~~~~(equal~(logbitp~0~x)~nil)))))} \\
\end{array}
\]

\subsection{Executability on Concrete Terms}
\label{sec:concrete-execution}

Suppose GL is symbolically executing a function call.  If the arguments to the
function are all concrete objects (i.e., symbolic objects that represent a
single value), then in some cases the interpreter can stop symbolically
executing and just run the ACL2 function on these arguments.  In some
cases, this can provide a critical performance boost.

To actually call these functions, GL essentially uses a case statement along
the following lines.

\[
\begin{array}{l}
\texttt{(case~fn} \\
\texttt{~~(cons~~~~~(cons~(first~args)~(second~args)))} \\
\texttt{~~(reverse~~(reverse~(first~args)))} \\
\texttt{~~(member~~~(member~(first~args)~(second~args)))} \\
\texttt{~~...)} \\
\end{array}
\]

Such a case statement is naturally limited to calling a fixed set of functions.
To allow GL to concretely execute additional functions, you can use
\texttt{def-gl-clause-processor}, a special macro that defines a new version of
the GL symbolic interpreter and clause processor.  GL automatically uses the
most recently defined interpreter and clause processor.  For instance, here is
the syntax for extending GL so that it can execute \texttt{md5sum} and
\texttt{sets::mergesort}:

\[
\texttt{(def-gl-clause-processor my-cp '(md5sum sets::mergesort))}.
\]

\subsection{Full-Custom Symbolic Counterparts}
\label{sec:custom-symbolic-counterparts}

% GL includes hand-written functions, called \emph{symbolic counterparts}, for
% symbolically executing primitive ACL2 functions like \texttt{+},
% \texttt{consp}, \texttt{equal}, and also important defined functions like
% \texttt{logand}, \texttt{ash}, etc.

The advanced GL user can write custom symbolic counterparts to get better
performance.  This is somewhat involved.  Generally, such a function operates
by cases on what kinds of symbolic objects it has been given.  Most of these
cases are easy; for instance, the symbolic counterpart for \texttt{consp} just
returns \texttt{nil} when given a \texttt{:g-boolean} or \texttt{:g-number}.
But in other cases the operation can require combining the Boolean
expressions making up the arguments in some way, e.g., the symbolic counterpart
for \texttt{binary-*} implements a simple binary multiplier.

Once the counterpart has been defined, it must be proven sound with respect to
the semantics of ACL2 and the symbolic object format.  This is an ordinary ACL2
proof effort that requires some understanding of GL's implementation.

The most sophisticated symbolic counterpart we have written is an AIG to BDD
conversion algorithm~\cite{10-swords-bddify}.  This function serves as a
symbolic counterpart for AIG evaluation, and at Centaur it is the basis for the
``implementation side'' of our hardware correctness theorems.  This algorithm
and its correctness proof are publicly available; see
\texttt{centaur/aig/g-aig-eval}.

\section{Case-Splitting}
\label{sec:def-gl-param-thm}

BDD performance can sometimes be improved by breaking a problem into subcases.
The standard example is floating-point
addition~\cite{98-chen-adders,99-aagaard-param}, which benefits from separating
the problem into cases based on the difference between the two inputs'
exponents.  For each exponent difference, the two mantissas are aligned
differently before being added together, so a different BDD order is necessary
to interleave their bits at the right offset.  Without case splitting, a single
BDD ordering has to be used for the whole problem; no matter what ordering we
choose, the mantissas will be poorly interleaved for some exponent differences,
causing severe performance problems.  Separating the cases allows the
appropriate order to be used for each difference.

GL provides a \texttt{def-gl-param-thm} command that supports this technique.
This command splits the goal formula into several subgoals and attempts to
prove each of them using the \texttt{def-gl-thm} approach, so for each subgoal
there is a symbolic execution step and coverage proof.  To show the subgoals
suffice to prove the goal formula, it also does another
\texttt{def-gl-thm}-style proof that establishes that any inputs satisfying the
hypothesis are covered by some case.

%  each of which is
% handled by a \texttt{def-gl-thm}-style proof that includes its own
% symbolic execution step and coverage proof.  To show the subgoals suffice to
% prove the goal formula, it also does another \texttt{def-gl-thm}-style proof
% establishing that any inputs satisfying the hypothesis are covered by some case.

Here is how we might split the proof
for \texttt{fast-logcount-32} into five subgoals.
  One goal handles the case
where the most significant bit is 1.  The other four goals assume the most
significant bit is 0, and separately handle the cases where the lower two bits
are 0, 1, 2, or 3.  Each case has a different symbolic binding for \texttt{x},
giving the BDD variable order. Of course, splitting into cases and varying the
BDD ordering is unnecessary for this theorem, but it illustrates how the
\texttt{def-gl-param-thm} command works.

\[
\begin{array}{l}
\texttt{(def-gl-param-thm~fast-logcount-32-correct-alt} \\
\texttt{~:hyp~(unsigned-byte-p~32~x)} \\
\texttt{~:concl~(equal~(fast-logcount-32~x)} \\
\texttt{~~~~~~~~~~~~~~~(logcount~x))} \\
\texttt{~:param-bindings} \\
\texttt{~`((((msb~1)~(low~nil))~((x~,(g-int~32~-1~33))))} \\
\texttt{~~~(((msb~0)~(low~0))~~~((x~,(g-int~~0~~1~33))))} \\
\texttt{~~~(((msb~0)~(low~1))~~~((x~,(g-int~~5~~1~33))))} \\
\texttt{~~~(((msb~0)~(low~2))~~~((x~,(g-int~~0~~2~33))))} \\
\texttt{~~~(((msb~0)~(low~3))~~~((x~,(g-int~~3~~1~33)))))} \\
\texttt{~:param-hyp~(and~(equal~msb~(ash~x~-31))} \\
\texttt{~~~~~~~~~~~~~~~~~(or~(equal~msb~1)} \\
\texttt{~~~~~~~~~~~~~~~~~~~~~(equal~(logand~x~3)~low)))} \\
\texttt{~:cov-bindings~`((x~,(g-int~0~1~33))))} \\
\end{array}
\]

We specify the five subgoals to consider using two new variables, \texttt{msb}
and \texttt{low}.  Here, \texttt{msb} will determine the most significant bit
of \texttt{x}; \texttt{low} will determine the two least significant bits of
\texttt{x}, but only when \texttt{msb} is 0.

The \texttt{:param-bindings} argument describes the five subgoals by assigning
different values to \texttt{msb} and \texttt{low}.  It also gives the
\texttt{g-bindings} to use in each case.  We use different bindings for
\texttt{x} for each subgoal to show how it is done.

%Here we use a different bindings for \texttt{x} for each subgoal
%just to show that this is possible, but in general different cases may require different
%symbolic bindings in order to obtain good performance.

The \texttt{:param-hyp} argument describes the relationship between
\texttt{msb}, \texttt{low}, and \texttt{x} that will be assumed in each
subgoal.  In the symbolic execution performed for each subgoal, the
\texttt{:param-hyp} is used to reduce the space of objects represented by the
symbolic binding for \texttt{x}.  For example, in the subgoal where
$\texttt{msb} = 1$, this process will assign \Btrue to $\texttt{x}[31]$.  The
\texttt{:param-hyp} will also be assumed to hold for the coverage proof for
each case.

How do we know the case-split is complete?  One final proof is needed to show
that whenever the hypothesis holds for some \texttt{x}, then at least one of
the settings of \texttt{msb} and \texttt{low} satisfies the \texttt{:param-hyp}
for this \texttt{x}.  That is:

\[
\begin{array}{l}
\texttt{(implies~(unsigned-byte-p~32~x)} \\
\texttt{~~~~~~~~~(or~(let~((msb~1)~(low~nil))} \\
\texttt{~~~~~~~~~~~~~~~(and~(equal~msb~(ash~x~-31))} \\
\texttt{~~~~~~~~~~~~~~~~~~~~(or~(equal~msb~1)} \\
\texttt{~~~~~~~~~~~~~~~~~~~~~~~~(equal~(logand~x~3)~low))))} \\
\texttt{~~~~~~~~~~~~~(let~((msb~0)~(low~0))~...)} \\
\texttt{~~~~~~~~~~~~~(let~((msb~0)~(low~1))~...)} \\
\texttt{~~~~~~~~~~~~~(let~((msb~0)~(low~2))~...)} \\
\texttt{~~~~~~~~~~~~~(let~((msb~0)~(low~3))~...)))} \\
\end{array}
\]

This proof is also done in the \texttt{def-gl-thm} style, so we need we need
one last set of symbolic bindings, which is provided by the
\texttt{:cov-bindings} argument.

\section{AIG Mode}
\label{sec:aig-mode}

GL can optionally use And-Inverter Graphs (AIGs) to represent
Boolean expressions instead of BDDs.  You can choose the
mode on a per-proof basis by running \texttt{(gl-bdd-mode)} or
\texttt{(gl-aig-mode)}, which generate \texttt{defattach} events.

Unlike BDDs, AIGs are non-canonical, and this affects performance in
fundamental ways.  AIGs are generally much cheaper to construct than BDDs, but
to determine whether AIGs are equivalent we have to invoke a SAT solver,
whereas with BDDs we just need to use a pointer-equality check.

Using an external SAT solver raises questions of trust.  For most verification
work in industry it is probably sufficient to just trust the solver.  But Matt
Kaufmann has developed and reflectively verified an ACL2 function that
efficiently checks a resolution proof that is produced by the SAT solver.  GL
can use this proof-checking capability to avoid trusting the SAT solver.  This
approach is not novel: Weber and Amjad~\cite{09-weber-sat} have developed an
LCF-style integration of SAT in several HOL theorem provers, and Darbari, et
al~\cite{10-darbari-sat} have a reflectively verified SAT certificate checker
in Coq.

Recording and checking resolution proofs imposes significant overhead, but is
still practical in many cases.  We measured this overhead on a collection of
AIG-mode GL theorems about Centaur's MMX/SSE module.  These theorems take 10
minutes without proof recording.  With proof-recording enabled, our SAT solver
uses a less-efficient CNF generation algorithm and SAT solving grows to 25
minutes; an additional 6 minutes are needed to check the recorded proofs.

The SAT solver we have been using, an integration of MiniSAT with an AIG package,
is not yet released, so AIG mode is not usable ``out of the box.''
As future work, we would like to make
it easier to plug in other SAT solvers.  Versions of MiniSAT, PicoSAT, and ZChaff can
also produce resolution proofs, so this is mainly an interfacing issue.

A convenient feature of AIGs is that you do not have to come up with a good
variable ordering.  This is especially beneficial if it avoids the need to
case-split.  On the other hand, BDDs provide especially nice counterexamples,
whereas SAT produces just one, essentially random counterexample.

Performance-wise, AIGs are better for some problems and BDDs for others.  Many
operations combine bits from data buses in a regular, orderly way; in these
cases, there is often a good BDD ordering and BDDs may be faster than SAT.
But when the operations are less regular, when no good BDD ordering is
apparent, or when case-splitting seems necessary to get good BDD performance,
SAT may do better.  For many of our proofs, SAT works well enough that we
haven't tried to find a good BDD ordering.

\section{Debugging Failures}
\label{sec:debugging}

A GL proof attempt can fail in several ways.  In the ``best'' case, the
conjecture is disproved and GL can produce counterexamples to help diagnose the
problem.  However, sometimes symbolic execution simply runs forever (Section
\ref{sec:performance-problems}).  In other cases, a symbolic execution may
produce an indeterminate result (Section \ref{sec:indeterminate-results}),
giving an example of inputs for which the symbolic execution failed.  Finally,
GL can run out of memory or spend too much time in garbage collection (Section
\ref{sec:memory-problems}).  We have developed some tools and techniques for
debugging these problems.

\subsection{Performance Problems}
\label{sec:performance-problems}

Any bit-blasting tool has capacity limitations.  However, you may also run into
cases where GL is performing poorly due to preventable issues.  When GL seems
to be running forever, it can be helpful to trace the symbolic interpreter to
see which functions are causing the problem.  To trace the symbolic
interpreter, run
\[
\texttt{(gl::trace-gl-interp~:show-values t)}.
\]
Here, at each call of the symbolic interpreter, the term being interpreted and
the variable bindings are shown, but since symbolic objects may be too large to
print, any bindings that are not concrete are hidden.  You can also get a trace
with no variable bindings using \texttt{:show-values nil}.  It may also be
helpful to simply interrupt the computation and look at the Lisp backtrace,
after executing
\[\texttt{(set-debugger-enable t)}.\]
In many cases, performance problems are due to BDDs growing too large.  This is
likely the case if the interpreter appears to get stuck (not printing any more
trace output) and the backtrace contains a lot of functions with
names beginning in \texttt{q-}, which is the convention for BDD operators.  In
some cases, these performance problems may be solved by choosing a more
efficient BDD order.  But note that certain operations like
multiplication are exponentially hard.  If you run into these
limits, you may need to refactor or decompose your problem into simpler
sub-problems (Section \ref{sec:def-gl-param-thm}).

There is one kind of BDD performance problem with a special solution.  Suppose
GL is asked to prove \texttt{(equal spec impl)} when this does not actually
hold.  Sometimes the symbolic objects for \texttt{spec} and \texttt{impl} can
be created, but the BDD representing their equality is too large to fit in
memory.  The goal may then be restated with \texttt{always-equal}
instead of \texttt{equal} as the final comparison.  Logically,
\texttt{always-equal} is just \texttt{equal}.  But \texttt{always-equal}
has a custom symbolic counterpart that returns \texttt{t} when its arguments
are equivalent, or else produces a symbolic object that captures just one
counterexample and is indeterminate in all other cases.

Another possible problem is that the symbolic interpreter never gets
stuck, but keeps opening up more and more functions.  These
problems might be due to redundant recursion (see Section
\ref{sec:redundant-recursion}), which may be avoided by providing a
more efficient preferred definition (Section \ref{sec:preferred-definitions})
for the function.  The symbolic interpreter might also be
inefficiently interpreting function calls on concrete arguments, in
which case a \texttt{def-gl-clause-processor} call may be used to allow GL
to execute the functions directly (Section \ref{sec:concrete-execution}).

\subsection{Indeterminate Results}
\label{sec:indeterminate-results}

Occasionally, GL will abort a proof and print a message saying it found
indeterminate results.  In this case, the examples printed are likely
\emph{not} to be true counterexamples, and examining them may not be
particularly useful.

One likely reason for such a failure is that some of GL's built-in symbolic counterparts
have limitations.  For example, most arithmetic primitives will
not perform symbolic computations on non-integer numbers.  When ``bad'' inputs are
provided, instead of producing a new \texttt{:g-number} object, these functions
will produce a \texttt{:g-apply} object, which is a type of symbolic
object that represents a function call.  A \texttt{:g-apply} object cannot be
syntactically analyzed in the way other symbolic objects can, so most
symbolic counterparts, given a \texttt{:g-apply} object, will simply create
another one wrapping its arguments.

To diagnose indeterminate results, it is helpful to know when the first
\texttt{:g-apply} object was created.  If you run
\[
\texttt{(gl::break-on-g-apply)},
\]
then when a \texttt{:g-apply} object is constructed, the function and symbolic
arguments will be printed and an interrupt will occur, allowing you to inspect
the backtrace.  For example, the following form produces an indeterminate result.
\[
\begin{array}{l}
\texttt{(def-gl-thm~integer-half} \\
\texttt{~~:hyp~(and~(unsigned-byte-p~4~x)} \\
\texttt{~~~~~~~~~~~~(not~(logbitp~0~x)))} \\
\texttt{~~:concl~(equal~(*~1/2~x)} \\
\texttt{~~~~~~~~~~~~~~~~(ash~x~-1))} \\
\texttt{~~:g-bindings~`((x~,(g-int~0~1~5))))} \\
\end{array}
\]
After running \texttt{(gl::break-on-g-apply)}, running the above form enters
a break after printing
\[
\texttt{(g-apply BINARY-* (1/2 (:G-NUMBER (NIL \# \# \# NIL)))}
\]
to signify that a \texttt{:g-apply} form was created after trying to multiply
some symbolic integer by $\frac{1}{2}$.

Another likely reason is that there is a typo in your theorem.  When a variable
is omitted from the \texttt{:g-bindings} form, a warning is printed and the
missing variable is assigned a \texttt{:g-var} object.  A \texttt{:g-var} can
represent any ACL2 object, without restriction.  Symbolic counterparts
typically produce \texttt{:g-apply} objects when called on \texttt{:g-var}
arguments, and this can easily lead to indeterminate results.

\subsection{Memory Problems}
\label{sec:memory-problems}

Memory management can play a significant role in symbolic execution
performance.  In some cases GL may use too much memory, leading to swapping and
slow performance.  In other cases, garbage collection may run too frequently or
may not reclaim much space.  We have several recommendations for managing
memory in large-scale GL proofs.  Some of these suggestions are specific to
Clozure Common Lisp.

1.  Load the \texttt{centaur/misc/memory-mgmt-raw} book and use the
\texttt{set-max-mem} command to indicate how large you would like the
Lisp heap to be.  For instance, \[ \texttt{(set-max-mem (* 8 (expt 2
  30)))} \] says to allocate 8 GB of memory.  To avoid swapping, you should
use somewhat less than your available physical memory.  This book disables
ephemeral garbage collection and configures the garbage collector to run only
when the threshold set above is exceeded, which can boost performance.

2.  Optimize hash-consing performance.  GL's representations of BDDs and AIGs
use \texttt{hons} for structure-sharing.  The \texttt{hons-summary} command can
be used at any time to see how many honses are currently in use, and
hash-consing performance can be improved by pre-allocating space for these
honses with \texttt{hons-resize}.  See the \texttt{:doc} topics for these
commands for more information.

3.  Be aware of (and control) hash-consing and memoization overhead.  Symbolic
execution can use a lot of hash conses and can populate the memoization tables
for various functions.  The memory used for these purposes is \emph{not}
automatically freed during garbage collection, so it may sometimes be necessary
to manually reclaim it.  A useful function is \texttt{(maybe-wash-memory~$n$)},
which frees this memory and triggers a garbage collection only when the amount
of free memory is below some threshold $n$.  A good choice for $n$ might be
20\% of the \texttt{set-max-mem} threshold.  It can be useful to call
\texttt{maybe-wash-memory} between proofs, or between the cases of
parametrized theorems; see \texttt{:doc def-gl-param-thm} for its
\texttt{:run-be\-fore-ca\-ses} argument.

\section{Related Work}
\label{sec:related}

GL is most closely related to Boyer and Hunt's~\cite{09-boyer-g} \emph{G}
system, which was used for earlier proofs about Centaur's floating-point unit.
G used a symbolic object format similar to GL's, but only supported BDDs.  It
also included a compiler that could produce ``generalized'' versions of
functions, similar to symbolic counterparts.  GL actually has such a compiler,
but the interpreter is more convenient since no compilation step is necessary,
and the performance difference is insignificant.  In experimental comparisons,
GL performed as well or better than G, perhaps due to the change from G's
sign/magnitude number encoding to GL's two's-complement encoding.

The G system was written ``outside the logic,'' in Common Lisp.  It could not
be reasoned about by ACL2, but an experimental connection was developed which
allowed ACL2 to trust G to prove theorems.  In contrast, GL is written entirely
in ACL2, and its proof procedure is a reflectively-verified clause processor,
which provides a significantly better story of trust.  Additionally, GL can be
safely configured and extended by users via preferred definitions and custom
symbolic counterparts.

Reeber~\cite{06-reeber-sulfa} identified a decidable subset of ACL2 formulas
called SULFA and developed a SAT-based procedure for proving theorems in this
subset.  Notably, this subset included lists of bits and recursive functions of
bounded depth.  The decision procedure for SULFA is not mechanically verified,
but Reeber's dissertation~\cite{07-reeber-dissertation} includes an argument
for its correctness.  GL addresses a different subset of ACL2 (e.g., SULFA
includes uninterpreted functions, whereas GL includes numbers and arithmetic
primitives), but the goals of both systems are similar.

ACL2 has a built-in BDD algorithm (described in \texttt{:doc bdd}) that, like
SULFA, basically deals with Booleans and lists of Booleans, but not numbers,
addition, etc.  This algorithm is tightly integrated with the prover; it can
treat provably Boolean terms as variables and can use unconditional rewrite
rules to simplify terms it encounters.  The algorithm is written in program
mode (outside the ACL2 logic) and has not been mechanically verified.  GL seems
to be significantly faster, at least on a simple series of
addition-commutativity theorems.

Fox~\cite{11-fox-blasting} has implemented a bit-blasting procedure in HOL4
that can use SAT to solve problems phrased in terms of a particular bit-vector
representation.  This tool is based on an LCF-style integrations of
proof-producing SAT solvers, so it has a strong soundness story.  We would
expect there to be some overhead for any LCF-style
solution~\cite{09-weber-sat}, and GL seems to be considerably faster on the
examples in Fox's paper; see the supporting materials for details.

Manolios and Srinivasan \cite{06-manolios-pipeline} describe a connection
between ACL2 and UCLID to verify that a pipelined processor implements its
instruction set.  In this work, ACL2 is used to simplify the correctness
theorem for a bit-accurate model of the processor down to a more abstract,
term-based goal.  This goal is then given to UCLID, a decision procedure for a
restricted logic of counter arithmetic, lambdas, and uninterpreted functions.
UCLID then proves the goal much more efficiently than, e.g., ACL2's rewriter.
This work seems complementary to GL, which deals with bit-level reasoning,
i.e., the parts of the problem that this strategy addresses using ACL2.

Srinivasan \cite{07-srinivasan-dissertation} additionally described ACL2-SMT, a
connection with the Yices SMT solver.  The system attempts to unroll and
simplify ACL2 formulas until they can be translated into the input language of
the SMT solver (essentially linear integer arithmetic, array operations, and
uninterpreted integer and Boolean functions).  It then calls Yices to discharge
the goal, and Yices is trusted.  GL addresses a different subset of ACL2, e.g.,
GL supports list operations and more arithmetic operations like
\texttt{logand}, but ACL2-SMT has uninterpreted functions and can deal with,
e.g., unbounded arithmetic.

Armand, et. al~\cite{11-armand-sat} describe work to connect SAT and SMT
solvers with Coq.  Unlike the ACL2-SMT work, the connection is carried out in a
verified way, with Coq being used to check proof witnesses generated by the
solvers.  This connection can be used to prove Coq goals that directly fit into
the supported logic of the SMT solver.  GL is somewhat different in that it
allows most any ACL2 term to be handled when its variables range over a finite
space.

% Other theorem provers (we already mentioned Fox's bit-blasting stuff, but
% more generally integrations of SMT with Coq, etc.)

% True bit-based stuff (ABC, CUDD, SMT solvers in general, ...)

\section{Conclusions}

GL provides a convenient and efficient way to solve many finite ACL2 theorems
that arise in hardware verification.  It allows properties to be stated in a
straightforward manner, scales to large problems, and provides clear
counter-examples for debugging.  At Centaur Technology, it
plays an important role in the verification of arithmetic units, and we make
frequent improvements to support new uses.

Beyond this paper, we encourage all GL users to see the online documentation,
which can be found under \texttt{:doc gl} after loading the GL library.  If you
prefer, you can also generate an HTML version of the documentation; see
\texttt{centaur/README} for details.  Finally, the documentation for ACL2(h)
may be useful, and can be found at \texttt{:doc hons-and-memoization}.

While we have described the basic idea of symbolic execution and how GL uses it
to prove theorems, Swords' dissertation~\cite{10-swords-dissertation} contains
a much more detailed description of GL's implementation.  It covers tricky
topics like the handling of \texttt{if} statements and the details of BDD
parametrization.  It also covers the logical foundations of GL, such as
correctness claims for symbolic counterparts, the introduction of symbolic
interpreters, and the definition and verification of the GL clause processor.

\subsection{Acknowledgments}

Bob Boyer and Warren Hunt developed the G system, which pioneered many of the
ideas in GL.  Anna Slo\-bo\-do\-v\'{a} has carried out several sophisticated
proofs with GL and beta-tested many GL features.  Matt Kaufmann and Niklas Een
have contributed to our verified SAT integration.  Gary Byers has answered many
of our questions and given us advice about Clozure Common Lisp.  We thank
Warren Hunt, Matt Kaufmann, David Rager, Anna Slo\-bo\-do\-v\'{a}, and the
anonymous reviewers for their corrections and feedback on this paper.

\bibliographystyle{eptcs}
\bibliography{paper}{}

\end{document}